\documentstyle[12pt]{article}
\begin{document}

\centerline{\bf Ambiguities in the partial-wave analysis of 
pseudoscalar-meson photoproduction}
\vskip .5cm
\centerline{Greg Keaton and Ron Workman}
\centerline{Department of Physics, Virginia Polytechnic Institute and State
University, Blacksburg, VA~24061}
 
\begin{abstract}
 
Ambiguities in pseudoscalar-meson photoproduction, 
arising from incomplete experimental data, 
have analogs in pion-nucleon scattering.
Amplitude ambiguities have important
implications for the problems of amplitude extraction and 
resonance identification in partial-wave
analysis. The effect of these ambiguities on observables is described. 
We compare our results with those found in earlier studies.   
 
\end{abstract}

\vskip .5cm
PACS Numbers: 13.60.Le, 13.88.+e
\eject

\centerline{I. Introduction}
\vskip .2cm

Our empirical knowledge of the $N$ (S=0, I=1/2) and $\Delta$ (S=0, I=3/2) 
baryons is mainly based upon data from the scattering and photoproduction of 
pseudoscalar mesons. Resonance positions and couplings have generally 
come from partial-wave analyses of the existing (incomplete) sets of
observables\cite{c1}. The lack of sufficient experimental data implies that 
the transition amplitudes cannot be uniquely determined. 
Barring further theoretical input, multiple sets of valid amplitudes exist. 

Typical analyses employ the additional constraints of unitarity and 
analyticity, which reduce the range of potential ambiguities\cite{amb}. 
Amplitudes
are expected to be `smooth' with the possible exception of threshold cusps.
Born terms are usually added, either diagrammatically or through the use 
of dispersion relations. The Carnegie-Mellon$-$Berkeley (CMB)\cite{cmb}
and Karlsruhe-Helsinki (KH)\cite{kh} groups used a wide range of 
dispersion relation constraints\cite{c2} in their analyses. As a result,
these independent studies produced results which were qualitatively 
very similar. However, recent spin-rotation data\cite{data} 
are in marked disagreement with the prediction of these analyses.
No data of this type were available (in the resonance region) when these
analyses were performed. 

We will concentrate on the ambiguities which can arise
in the partial-wave analysis of pseudoscalar meson photoproduction data.
There is a close analogy between the ambiguities found in the photoproduction
and the elastic scattering of pseudoscalar mesons. This is particularly
evident if one adopts the method of Dean and Lee\cite{dean}. In a previous
work\cite{kw} we considered the problems encountered in constructing a 
complete experiment. The present study is more general. Here we will show
how amplitude ambiguities can alter the angular structure of observables and
these results will be compared to some 
earlier findings of Omelaenko\cite{Omel}. 
We will also mention how these results are related to the study of 
nodal trajectories\cite{tab}.

\vskip .2cm
\centerline{II. Conjugation Symmetries}
\vskip .2cm

As suggested in the Introduction, the ambiguities associated with 
pseudoscalar meson photoproduction are most easily described in analogy
with elastic meson-nucleon scattering. To that end, we will first define the
elastic scattering amplitudes.  Following the notation of Ref.\cite{dean},
the transition matrix is given by
\begin{equation}
T \; = \; F \; + \; iG \hat{n} \cdot \vec{\sigma} ,
\end{equation}                 
where $\hat{n}$ is the normal to the scattering plane. 
The spin-flip ($G$) and non-flip ($F$) amplitudes can be decomposed into
partial-wave amplitudes
\begin{eqnarray}
F(\theta ) &=& \sum_l [ (l+1) f_{l+} + l f_{l-} ] 
P_l (\cos \theta ) , \\
G(\theta ) &=& \sum_l (f_{l+} - f_{l-} )
\sin \theta P_l ' (\cos \theta ) ,
\end{eqnarray}
where the subscript $l\pm$ gives the $J$-value, $J=l\pm 1/2$, and
$\theta$ is the center-of-mass scattering angle. 

In terms of these amplitudes, the differential cross section 
($d\sigma / d\Omega$) and polarization ($P$) are 
\begin{eqnarray}
{d\sigma \over {d\Omega} } &=& |F|^2 + |G|^2 ,   \\
P {d\sigma \over {d\Omega} } &=& -2 {\rm Im} F^* G .
\end{eqnarray}

We will first consider a transformation 
\begin{equation}
\left( \begin{array}{c} F \\ G \end{array} \right) \to  
\left( \begin{array}{c} -F^* \\ G^* \end{array} \right) ,  
\end{equation}
which preserves both the cross section and polarization.  
Therefore experimental information on the differential cross section and 
polarization alone are insufficient to determine $F$ and $G$.

The photoproduction amplitude can be similarly divided into spin single-flip
($S_1$, $S_2$), spin non-flip ($N$), and spin double-flip ($D$) 
pieces\cite{Bark}. A transformation analogous to Eq.(6) is Ambiguity IV 
of Ref.\cite{kw} :              
\begin{equation}
\left( \begin{array}{c} S_1 \\ S_2 \end{array} \right) \to  
\left( \begin{array}{c} -S_1^* \\ -S_2^* \end{array} \right)  
\;\; {\rm and} \;\;
\left( \begin{array}{c} N \\ D \end{array} \right) \to  
\left( \begin{array}{c} N^*  \\ D^* \end{array} \right) , 
\end{equation}
which is a symmetry of the cross section,
single-polarization observables, and half of the double-polarization
observables listed in Ref.\cite{Bark}.

The associated change in partial-wave amplitudes is clear if we first 
introduce the helicity amplitudes and helicity elements of
Walker\cite{walker}
\begin{eqnarray}
S_1 &=& {1\over \sqrt{2} } \sin \theta \cos {1\over 2} \theta
\sum_{l=1}^{\infty} (B_{l+} - B_{(l+1)-}) (P_l '' - P_{l+1} '') ,\\
D &=& {1\over \sqrt{2} } \sin \theta \sin {1\over 2} \theta
\sum_{l=1}^{\infty} (B_{l+} + B_{(l+1)-}) (P_l '' + P_{l+1} '') ,\\
N &=& \sqrt{2}  \cos {1\over 2} \theta
\sum_{l=0}^{\infty} (A_{l+} - A_{(l+1)-}) (P_l ' - P_{l+1} ') ,\\
S_2 &=& \sqrt{2}  \sin {1\over 2} \theta
\sum_{l=0}^{\infty} (A_{l+} + A_{(l+1)-}) (P_l ' + P_{l+1} ') .
\end{eqnarray}
The transformation given in Eq.(7) is then equivalent to an 
exchange of helicity elements
\begin{equation}
B_{l+} \leftrightarrow B_{(l+1)-}^*  \;\; {\rm and} \;\;
A_{l+} \leftrightarrow -A_{(l+1)-}^* .
\end{equation}
It should be noted that this transformation is only pertinent above the 
$\pi \pi N$ threshold. At lower energies it violates unitarity in the form of
Watson's theorem\cite{watson}.

\vskip .2cm
\centerline{III. Continuous Symmetries}
\vskip .2cm
					   
As discussed in Ref.\cite{dean}, the polarization and cross section 
for elastic scattering are also invariant under rotations of the 
$F$ and $G$ amplitudes
\begin{equation}
\left( \begin{array}{c}
	     F' \\ G'  \end{array} \right)   = 
\left( \begin{array}{cc} \cos \phi & \sin \phi \\             
			-\sin \phi & \cos \phi \end{array} \right)
\left( \begin{array}{c}
	     F \\ G \end{array} \right) .
\end{equation}
Here $\phi$ is a parameter which can vary with the energy
and scattering angle. While this transformation does not preserve elastic
unitarity, it has implications for resonance identification above the
inelastic threshold\cite{dean}. As noted in Ref.\cite{dean}, if this
rotation (with $\phi = - \theta$) 
is composed with the conjugation operation given in Eq.(6),
the Minami ambiguity\cite{min} 
\begin{equation}
f_{l\pm} \; \to \; -f^*_{ (l\pm 1)\mp } ,
\end{equation}
results. This transformation, applied to the partial-wave amplitudes,
preserves elastic unitarity along with the cross section and polarization.

The above rotation also has an analog in terms of   
photoproduction amplitudes.
For example, Ambiguity III of Ref.\cite{kw} is given by 
\begin{equation}
\left( \begin{array}{c} S_1 \\ D \end{array} \right) \to  
\left( \begin{array}{c} D \\ -S_1 \end{array} \right)  
\;\; {\rm and} \;\;
\left( \begin{array}{c} N \\ S_2 \end{array} \right) \to  
\left( \begin{array}{c} S_2  \\ -N \end{array} \right) , 
\end{equation}
which is a special case ($\phi=\pi /2$) of the more general 
transformation 
\begin{equation}
\left( \begin{array}{c}
	    S_1 ' \\ D' \\ N ' \\ S_2' \end{array} \right) =
\left( \begin{array}{cccc} \cos \phi & \sin \phi & 0 & 0 \\            
			 - \sin \phi & \cos \phi & 0 & 0 \\
				 0   &  0   & \cos \phi & \sin \phi \\
				 0   &  0   & -\sin \phi & \cos \phi 
				 \end{array} \right)
\left( \begin{array}{c}
	    S_1 \\ D \\ N \\ S_2 \end{array} \right)  .
\end{equation}
Here also $\phi$ depends on the energy and scattering angle. 
Ambiguities I and II of Ref. \cite{kw} can be generalized
in a similar way.  While a constant value of $\phi$ was chosen in 
Ref.\cite{kw}, the choice  $\phi = \phi (\theta)$ is more interesting.  
(In fact, $\phi$ {\it must} vary with the scattering angle $\theta$ 
\cite{c3}.)  The simplest choice, $\phi = \theta$, was shown\cite{c4} 
to confuse the identification of resonances in elastic scattering. 
The choice\cite{dean} $\phi = \epsilon \sin \theta$, for a small 
(angle-independent) parameter
$\epsilon$, is also interesting as it illustrates a case where 
solutions may be continuously varied with $\epsilon$. 

The cross section, single-polarization, and beam-target double-polarization
observables are invariant under the above transformation. The beam-recoil 
and target-recoil observables are not. 

\vskip .2cm
\centerline{IV. Fitting Angular Distributions}
\vskip .2cm

So far, we have not explicitly considered the problems which arise in
fitting angular distributions. Here one generally adopts the methods
of Barrelet\cite{Barrelet} or Gersten\cite{Gersten} in order to write
the transversity amplitudes as factorized polynomials in some function 
of the scattering angle. The case of $\pi N$ elastic scattering has been 
reviewed by H\"ohler\cite{kh}. Here we will concentrate on 
photoproduction, following the treatment given by Omelaenko\cite{Omel}. 

The use of transversity amplitudes 
\begin{eqnarray}
b_1 &=& {1\over 2} \left[ (S_1 + S_2) \; + \; i(N-D)\right] , \\ 
b_2 &=& {1\over 2} \left[ (S_1 + S_2) \; - \; i(N-D)\right] , \\ 
b_3 &=& {1\over 2} \left[ (S_1 - S_2) \; - \; i(N+D)\right] , \\ 
b_4 &=& {1\over 2} \left[ (S_1 - S_2) \; + \; i(N+D)\right] , 
\end{eqnarray}
allows the problem to be stated very simply. Measurements of the differential
cross section and single-polarization observables determine only the moduli
of $b_1$ through $b_4$, not their phases.  This leaves four undetermined 
phases.  However one overall phase is not observable, so there remain three
unknown phases.  These three unknowns correspond to the first three ambiguities
of Ref. \cite{kw}, which when expressed in the $b_i$-basis and generalized for
arbitrary angle $\phi$ as in Eq. (16) become
\[  
{\rm I}: \;\;\;\;
\left( \begin{array}{c}
	    b_1 ' \\ b_2 ' \\ b_3 ' \\ b_4 ' \end{array} \right) =
\left( \begin{array}{cccc} e^{-i\phi} & 0 & 0 & 0 \\            
			 0 & e^{-i\phi} & 0 & 0 \\
				 0   &  0   & e^{i\phi} & 0 \\
				 0   &  0   & 0 & e^{i\phi} 
				 \end{array} \right)
\left( \begin{array}{c}
	    b_1 \\ b_2 \\ b_3 \\ b_4 \end{array} \right)  ,
\]
\[ 
{\rm II}: \;\;\;\;
\left( \begin{array}{c}
	    b_1 ' \\ b_2 ' \\ b_3 ' \\ b_4 ' \end{array} \right) =
\left( \begin{array}{cccc} e^{-i\phi} & 0 & 0 & 0 \\            
			 0 & e^{i\phi} & 0 & 0 \\
				 0   &  0   & e^{i\phi} & 0 \\
				 0   &  0   & 0 & e^{-i\phi} 
				 \end{array} \right)
\left( \begin{array}{c}
	    b_1 \\ b_2 \\ b_3 \\ b_4 \end{array} \right)  ,
\]
\begin{equation}
{\rm III}: \;\;\;\;
\left( \begin{array}{c}
	    b_1 ' \\ b_2 ' \\ b_3 ' \\ b_4 ' \end{array} \right) =
\left( \begin{array}{cccc} e^{i\phi} & 0 & 0 & 0 \\            
			 0 & e^{-i\phi} & 0 & 0 \\
				 0   &  0   & e^{i\phi} & 0 \\
				 0   &  0   & 0 & e^{-i\phi} 
				 \end{array} \right)
\left( \begin{array}{c}
	    b_1 \\ b_2 \\ b_3 \\ b_4 \end{array} \right)  .
\end{equation}
Since the differential cross-section and single polarization observables
give no information about the phases of $b_1$ through $b_4$, 
it would appear that
the angles $\phi$ above are completely arbitrary.  However, this is not
so.  The form of the multipole expansion \cite{Omel} requires that
\begin{equation}
b_1 (\theta ) = - b_2 ( - \theta ) \;\; {\rm and} \;\; 
b_3 (\theta ) = - b_4 ( - \theta ) ,
\end{equation}
which restricts the dependence of $\phi$ on $\theta$. In ambiguity I,
$\phi$ must be an even function of $\theta$ while in ambiguities II 
and III, $\phi$ must be an odd function of $\theta$.  

The constraint given in Eq.(22) allowed Omelaenko\cite{Omel} to parameterize
the four transversity amplitudes in terms of two functions\cite{Gersten}
\begin{eqnarray}
b_1 &=& c a_{2L} {{ e^{i\theta / 2} }\over (1+x^2)^L }
\prod_{i=1}^{2L} ( x - \alpha_i ) \; , \\
b_3 &=& - c a_{2L} {{ e^{i\theta / 2} }\over (1+x^2)^L }
\prod_{i=1}^{2L} ( x - \beta_i ) \; , 
\end{eqnarray}
with $x = \tan \theta / 2$. 
Ambiguities result from the fact that complex conjugation of the
roots ($\alpha_i$ and $\beta_i$) alters the relative phases 
(but not the moduli) of the transversity amplitudes. 
One further condition
\begin{equation}
\prod_{i=1}^{2L} \alpha_i \; = \; \prod_{i=1}^{2L} \beta_i
\end{equation}
restricts this freedom. The most simple case (all roots conjugated)
is equivalent to the composition of the two transformations IV (Eq. (7)) 
and III with $\phi = \theta$.   
The choice of a finite cutoff in $L$ further restricts the 
values of $\phi (\theta)$ appearing in Eq.(21).  This is because if,
for example, $b_1' = b_1 e^{i\theta}$, and $b_1'$ is to be re-expressed
in the form of Eq. (23), the product must go to $i = 2(L+1)$.  Therefore
if the product is restricted to $i=2L$, this transformation
is ruled out.  In this case, the only indeterminacy is 
the freedom to conjugate the roots.

The effect of root conjugation was demonstrated in 
Fig. 5 of Ref.\cite{Omel}. Some of the double-polarization 
observables changed dramatically. 
This transformation was originally applied, however, to 
pion photoproduction amplitudes in the 
first resonance region, where Watson's theorem and the threshold
energy dependence can be used to resolve the ambiguities.  It would
be interesting to examine the effect of the ambiguities at higher
energies, where such constraints do not exist.
The region with a
center of mass energy near 1.9 GeV seems promising. This is the energy
at which the recent ITEP-PNPI spin-rotation measurements were made. 
Here there are many overlapping resonance candidates and we are well 
separated from the threshold region. 

We should also mention a recent study where 
these ambiguities could have important consequences. 
The nodal trajectory method\cite{tab} is concerned with the number (and 
energy dependence) of nodes 
found in photoproduction observables. Observables are split into 
`Legendre classes' having similar nodal structure.  
However, this grouping of observables is not respected by the transformations
we have discussed. 

As a test case, we chose 
the target-recoil observable $L_Z$ for $\gamma p\to p\pi^0$. 
Helicity amplitudes were 
generated from a multipole analysis\cite{said}, and $L_Z$ crossed zero  
three times at 500 MeV. Then the transformation given in Eq.(16)
was applied with 
$\phi = n\theta$. Using  $\phi$ = $\theta$ and 2$\theta$, the number of zero 
crossings increased to 5 and 7 respectively. The work of Omelaenko\cite{Omel}
indicates that the nodal structure can also be altered by the 
(smaller) set of ambiguities remaining when a fixed and finite 
angular momentum cutoff is applied. 
At sufficiently low energies, a knowledge of the 
threshold energy-dependence helps to resolve ambiguities. 
At higher energies, further assumptions seem necessary\cite{c6}. 

For the photoproduction of kaons and etas the problem is more acute. In 
analyzing these reactions, we have no Watson's theorem constraint and we
must account for the effect of sub-threshold resonances. It should also be 
noted that, in analyzing pion photoproduction data, the resonance positions
are usually taken as known from elastic $\pi N$ analyses. Given the 
possibility of significant contributions from `missing resonances' (that is,
resonances very weakly coupled to $\pi N$), kaon and eta photoproduction
analyses are relatively free of {\it a priori} constraints.  Therefore
they are more likely to be plagued by the kind of ambiguity discussed here.

\vskip .2cm
\centerline{V. Summary and Conclusions} 
\vskip .2cm

Pion photoproduction amplitudes are not completely determined by cross-section
and single polarization measurements.  This fact is exhibited by the 
existence of one discrete (Eq. (7)) and three continuous (Eq. (21))
transformations of the amplitudes that leave these observables invariant.
The transformations, introduced in Ref. \cite{kw}, are generalized in
this paper.  We have also shown how these transformations are related to the
ambiguity found by Omelaenko \cite{Omel}.

In order to resolve these ambiguities, either further data or more theoretical
input must be used.  One theoretical constraint comes from restricting the
amplitudes to contain only a certain number of partial waves.  As shown in
section IV, this reduces the ambiguities involved.  However, such a
theoretical restriction seems artificial, and cannot be justified in
the case of charged-pion photo-production (due to the $t$-channel pole).

Other constraints come from unitarity and the elastic $\pi N$ scattering data.
For energies between the $\pi N$ and $\pi\pi N$ thresholds, Watson's theorem
gives the phases of the photoproduction multipoles in terms of the elastic
$\pi N$ phase shifts.  This greatly reduces the ambiguity in the photoproduction
amplitudes.  Above the $\pi\pi N$ threshold such a powerful constraint
does not exist.  However, $\pi N$ data can again be used to reduce the
ambiguity.  We know the masses, widths, and $\pi N$ couplings of the
dominant resonances in the $\pi N$ channel 
(such as the $P_{33}(1232)$, $D_{13}(1520)$, and $F_{15}(1680)$ ).  
We can reject any transformation
of the photoproduction amplitudes that significantly alters these
parameters. Unfortunately, less is known about the resonances contributing 
to eta and kaon photoproduction.

The ambiguities discussed here are more relevant at higher energies,
where there are fewer theoretical restrictions, than at lower energies,
where Watson's theorem applies.  This has an important consequence for
the nodal trajectory method \cite{tab}, since the Legendre classes it
employs are not respected by the ambiguity transformations.  Therefore,
at energies where these transformations are allowed, the nodal trajectory
method will have to account for this additional freedom.

\vskip .2cm

This work was supported in part by the U.S. Department of
Energy Grants DE-FG05-88ER40454 and DE-FG05-95ER40709A.


\begin{thebibliography}{99}
\bibitem{c1} It is often said that $2N-1$ carefully chosen observables
	     are required in order to reconstruct $N$ independent 
	     amplitudes, up to an overall phase. 
	     That this is not generally true (due to quadrant ambiguities) 
	     is certainly `long known' if not `well known'. See, for 
	     example, Ref.\cite{Bark}.
\bibitem{c3} Strictly speaking, the choice of a constant $\phi$-value for
	     all values of $\theta$ is inappropriate. This is because, for
	     example, if $\phi = \pi/2$ in Eq. (16), then the
	     observable $C_z$ will change sign at all angles, and yet
	     it is constrained to equal $-1$ at $\theta = 0$ and +1
	     at 180 degrees.  Choices such as $\phi = \theta$ or 
	     $\phi = \epsilon \sin \theta$ 
	     leave all observables invariant at 0 and 180 degrees. 
             Some choices, such as $\phi = \theta$, can be eliminated
             if we require that the amplitudes smoothly connect to
             energy regions where this ambiguity is resolved.
\bibitem{c4} In Ref.\cite{dean} the effect of choosing $\phi = \theta$
	     was demonstrated using the partial-wave expansions for $F$ 
	     and $G$, along with recursion relations for the Legendre 
	     polynomials. In photoproduction, a similar result follows from 
	     the recursion relations for rotation functions,
	     using $\sin \theta = - \sqrt{2} d^1_{10}$ and 
	     $\cos \theta = d^1_{00}$,
	     along with the expansion for 
	     $d^J_{\lambda \mu} d^{J'}_{\lambda ' \mu '}$ (see for example, 
	     A.R. Edmonds, {\it Angular Momentum in Quantum Mechanics}
	     (Princeton, New Jersey, 1957), Section 4.3).
	     However, since this choice is motivated by simplicity 
	     (and not physics), we have omitted the details.
\bibitem{c6}  In a typical energy-dependent analysis, these ambiguities
	      are implicitly removed through a restriction of the analytic 
	      forms used in fitting the data.
\end{thebibliography}
\end{document}